\documentclass[twocolumn,trackchanges]{aastex62}

\received{}
\revised{}
\accepted{}
\submitjournal{ApJ}

\shorttitle{Comparative Study of Data-driven Coronal Field Models}
\shortauthors{Toriumi et al.}

\begin{document}

\title{Comparative Study of Data-driven Solar Coronal Field Models Using a Flux Emergence Simulation as a Ground-truth Data Set}

\correspondingauthor{Shin Toriumi}
\email{toriumi.shin@jaxa.jp}

\author[0000-0002-1276-2403]{Shin Toriumi}
\affiliation{Institute of Space and Astronautical Science (ISAS)/Japan Aerospace Exploration Agency (JAXA), 3-1-1 Yoshinodai, Chuo-ku, Sagamihara, Kanagawa 252-5210, Japan}

\author[0000-0003-3882-3945]{Shinsuke Takasao}
\affiliation{National Astronomical Observatory of Japan, 2-21-1 Osawa, Mitaka, Tokyo 181-8588, Japan}

\author[0000-0003-2110-9753]{Mark C.M. Cheung}
\affiliation{Lockheed Martin Solar and Astrophysics Laboratory, 3251 Hanover Street, Building/252, Palo Alto, CA 94304, USA}
\affiliation{Stanford University, Stanford, CA, USA}

\author[0000-0002-7018-6862]{Chaowei Jiang}
\affiliation{Institute of Space Science and Applied Technology, Harbin Institute of Technology, Shenzhen 518055, People's Republic of China}
\affiliation{SIGMA Weather Group, State Key Laboratory for Space Weather, National Space Science Center, Chinese Academy of Sciences, Beijing 100190, People's Republic of China}

\author[0000-0002-9293-8439]{Yang Guo}
\affiliation{School of Astronomy and Space Science and Key Laboratory for Modern Astronomy and Astrophysics, Nanjing University, Nanjing 210023, People's Republic of China}

\author[0000-0001-9046-6688]{Keiji Hayashi}
\affiliation{NorthWest Research Associates, Boulder, CO 80301, USA}
\affiliation{Stanford University, Stanford, CA, USA}

\author{Satoshi Inoue}
\affiliation{Institute for Space-Earth Environmental Research (ISEE), Nagoya University, Furo-cho, Chikusa-ku, Nagoya 464-8601, Japan}



\begin{abstract}
For a better understanding of magnetic field in the solar corona and dynamic activities such as flares and coronal mass ejections, it is crucial to measure the time-evolving coronal field and accurately estimate the magnetic energy. Recently, a new modeling technique called the data-driven coronal field model, in which the time evolution of magnetic field is driven by a sequence of photospheric magnetic and velocity field maps, has been developed and revealed the dynamics of flare-productive active regions. Here we report on the first qualitative and quantitative assessment of different data-driven models using a magnetic flux emergence simulation as a ground-truth (GT) data set. We compare the GT field with those reconstructed from the GT photospheric field by four data-driven algorithms. It is found that, at least, the flux rope structure is reproduced in all coronal field models. Quantitatively, however, the results show a certain degree of model dependence. In most cases, the magnetic energies and relative magnetic helicity are comparable to or at most twice of the GT values. The reproduced flux ropes have a sigmoidal shape (consistent with GT) of various sizes, a vertically-standing magnetic torus, or a packed structure. The observed discrepancies can be attributed to the highly non-force-free input photospheric field, from which the coronal field is reconstructed, and to the modeling constraints such as the treatment of background atmosphere, the bottom boundary setting, and the spatial resolution.
\end{abstract}

\keywords{Magnetohydrodynamics --- Solar active regions --- Solar corona --- Solar photosphere --- Solar magnetic fields}


\section{Introduction} \label{sec:intro}

Magnetic field plays a central role in driving a broad spectrum of energy-releasing activities of the Sun, most prominently represented by the solar flares and coronal mass ejections (CMEs) emanating from strongly magnetized active regions \citep{2011LRSP....8....6S,2011LRSP....8....1C,2019LRSP...16....3T}. Therefore, in order to better understand the nature of these magnetohydrodynamic (MHD) phenomena, it is vital to measure the magnetic field, track its evolution, and accurately estimate the storage and release of magnetic energy in a three-dimensional (3D) domain.

As of now, however, most of the magnetic field measurements are made at the photosphere and, albeit not that common, on the chromosphere. Because the coronal field is weak and the Doppler width is broad, it is not easy to resolve the Zeeman splitting of coronal emission lines and, therefore, there have been very few direct measurements of the coronal field \citep[e.g.,][]{2004ApJ...613L.177L}. Thanks to the frozen-in condition of plasma, one may trace the coronal loops in EUV and X-ray images and speculate the field configuration. Yet, complex magnetic structures that reside in flare-productive active regions may not be disentangled merely by morphological assessment of coronal imaging \citep{2014ApJ...785...34A}.

One possible solution to this problem is to employ extrapolation of magnetic fields from the observed photospheric magnetic field maps. A popular coronal field model assumes that: the magnetic field \mbox{\boldmath $B$} is static; all non-magnetic forces such as gas pressure gradient and gravity are negligible; and thus the Lorentz force vanishes. This assumption is called the force-free approximation and described as
\begin{eqnarray}
  \mbox{\boldmath $j$}\times \mbox{\boldmath $B$}=0,
  \label{eq:force-free}
\end{eqnarray}
where $\mbox{\boldmath $j$}=(c/4\pi)\nabla\times\mbox{\boldmath $B$}$ is the electric current density and $c$ is the speed of light. Equation (\ref{eq:force-free}) states that the electric current is parallel to the magnetic field everywhere in the coronal volume under consideration. Often, this condition is further rewritten as
\begin{eqnarray}
  \nabla\times\mbox{\boldmath $B$}=\alpha\mbox{\boldmath $B$},
  \label{eq:alpha}
\end{eqnarray}
which is mathematically equivalent to the Beltrami field of fluid dynamics. If $\alpha=0$, or equivalently $\mbox{\boldmath $j$}=0$ (current-free), the coronal field is called the potential field (PF), where the magnetic field is in the minimum energy state. The magnetic field is called linear force-free field (LFFF) if $\alpha$ is non-zero and uniform throughout the domain; otherwise, non-linear force-free field (NLFFF). Here, $\nabla\cdot\mbox{\boldmath $B$}=0$ requires $\alpha= {\rm const.}$ on the field lines and hence the non-linearity. In addition, the LFFF is the minimum energy state that conserves magnetic helicity.

In the last decade, in pace with the increasing availability of vector magnetograms delivered in particular by {\it Hinode} \citep{2007SoPh..243....3K} and {\it Solar Dynamics Observatory} \citep{2012SoPh..275....3P}, the NLFFF modeling has gained broad attention since energetic flares and CMEs emanate from active regions that harbor a large degree of non-potentiality \citep[][and references therein]{2019LRSP...16....3T,2019ApJ...886L..21T}. Although the model validations by \citet{2006SoPh..235..161S} and \citet{2009ApJ...696.1780D} revealed that different NLFFF schemes do not necessarily reproduce a consistent coronal field, these methods have been applied to a variety of targets \citep[see][for further reviews]{2012LRSP....9....5W,2016PEPS....3...19I,2017ScChE..60.1408G}.

However, because the eruptive phenomena are highly dynamic, the time series of static extrapolations does not capture accelerating states, and this is why the implementation of time-evolving coronal field models is now rapidly growing. One of these new approaches is the data-constrained model. In this concept, as an initial condition, the 3D coronal field is first constructed from a snapshot magnetogram using extrapolations such as the NLFFF technique. The dynamical evolution of the coronal field is then achieved by solving time-developing equations. Several authors have applied this type of model to the eruptive events. They triggered the unstable evolution by imposing velocity perturbation \citep{2012ApJ...758..117Z,2014Natur.514..465A}, inserting emerging flux \citep{2011ApJ...740...68F,2017ApJ...842...86M}, introducing anomalous resistivity \citep{2018NatCo...9..174I}, or simply by residual Lorentz force in the initial extrapolation fields, etc. \citep{2013ApJ...771L..30J,2013ApJ...779..129K}.

An even more realistic and advanced, but as yet immature, methodology is the data-driven model, in which the coronal field evolves in response to the sequentially updated photospheric boundary. Often the set of observed photospheric magnetic field and velocity field maps is used to advance the model coronal field forward in time. To date, multiple data-driven algorithms have been proposed, the main differences arising from the choice of governing equations (MHD vs. magnetofrictional), implementation of input bottom boundary (magnetic field-driven vs. electric field-driven), treatment of background atmosphere (stratified vs. uniform vs. zero-$\beta$\footnote{Plasma-$\beta$ is defined as the ratio between the gas pressure $p$ and the magnetic pressure $B^{2}/8\pi$: $\beta=p/(B^{2}/8\pi)$. Since the plasma-$\beta$ is generally very small in the corona ($\beta\ll 1$), the zero-$\beta$ approximation, in which the gas pressure gradient and gravity are neglected, is often employed.}), etc. These models have been applied to individual active regions and successfully reproduced the observed coronal loop structures and the resultant flares and CMEs \citep[e.g.,][]{2012ApJ...757..147C,2016NatCo...711522J,2016ApJ...828...62J,2017ApJ...838..113L,2018ApJ...855...11H,2019ApJ...871L..28H,2019ApJ...870L..21G,2019SoPh..294...41P,2019A&A...626A..91L}.

Although these studies demonstrate that the data-driven model is in fact a powerful tool to investigate the evolution of coronal field, it is still difficult to find out whether these algorithms reproduce the actual magnetic structure. Strengths and weaknesses of distinct models cannot be easily examined by independent analysis. In addition, uncertainties of observationally obtained vector magnetic fields such as noise and the 180$^{\circ}$ ambiguity \citep[e.g.,][]{2006SoPh..237..267M} inhibit the critical assessment of modeling accuracy using solar data.

Therefore, we set the primary goal of this study to quantitatively compare different types of data-driven coronal field models on common ground and understand their characteristics. To this end, we employ a 3D MHD simulation of magnetic flux emergence from the solar interior all the way up into the corona as a ground-truth (GT) data set, and examine the models' abilities to reproduce the GT coronal field. More precisely, we distribute to individual data-driven algorithms a sequence of photospheric slices of magnetic and velocity fields of the flux emergence simulation. Then, for each algorithm, we reconstruct the coronal field by adopting these synthetic observables as the series of bottom boundary. Finally, we compare the GT field with those reconstructed from different modeling algorithms.

The flux emergence simulations show a dynamical, non-force-free evolution across multiple atmospheric layers \citep{2009LRSP....6....4F,2014LRSP...11....3C}. Therefore, these simulations are suitable for evaluating how well the data-driven models, which often assume simple background atmospheres, can reproduce the GT magnetic field \citep{2017ApJ...838..113L}. In addition, an issue of observational uncertainties is naturally circumvented in this approach.

It should be emphasized here that our aim is not to rank the data-driven models. Our trial is just one example and other examinations may lead to different results. Nevertheless, we believe that the uniqueness of this research lies in the fact that this is the very first attempt to systematically survey various types of data-driven models.

The rest of this paper proceeds as follows. Section \ref{sec:model} describes the models that we investigate and Section \ref{sec:results} shows the results of the model comparison. Then, we summarize and discuss the results in Section \ref{sec:discussion}.

\section{Model Description} \label{sec:model}

In this work, we compared the GT flux emergence simulation with four different time-dependent data-driven coronal field models. In this section, we show the numerical settings of the GT simulation and each of the four reconstruction models. A brief overview of these models is provided as Table \ref{tab:model}. In addition, in order for an extensive comparison, we also reproduced the coronal field by using a NLFFF extrapolation code, which is shown in Appendix \ref{app:nlfff}.

\begin{deluxetable*}{cllccl}
\tablecaption{Summary of the numerical models\label{tab:model}}
\tablewidth{0pt}
\tablehead{
\colhead{Model} & \colhead{Type\tablenotemark{a}} & \colhead{Atmosphere} &
\colhead{Box size\tablenotemark{b}} &
\colhead{Grid number\tablenotemark{c}} & \colhead{References}}
\startdata
GT & self-consistent MHD & stratified & $600\times 600\times 540$ & $720\times 720\times 650$ & \citet{2017ApJ...850...39T}\\
MF & E-driven MF & N/A & $330\times 330\times 330$ & $400\times 400\times 400$ & \citet{2012ApJ...757..147C}\\
MHD1 & B-driven MHD & stratified & $846\times 846\times 846$ & $512\times 512\times 512$ & \citet{2016NatCo...711522J,2016ApJ...828...62J}\\
MHD2 & B-driven MHD & stratified & $330\times 330\times 330$ & $400\times 400\times 400$ & \citet{2019ApJ...870L..21G}\\
MHD3 & E-driven MHD & uniform & $330\times 330\times 330$ & $200\times 200\times 200$ & \citet{2018ApJ...855...11H,2019ApJ...871L..28H}\\
\enddata
\tablenotetext{a}{For the data-driven models, the input photospheric boundary condition is indicated by ``B-driven'' (magnetic field-driven) and ``E-driven'' (electric field-driven).}
\tablenotetext{b}{In the unit of $H_{0}\,(=170\ {\rm km})$.}
\tablenotetext{c}{GT has non-uniform grids, but the photospheric slices are converted to uniform spacing before distributed to each coronal field model. All other models have uniform grids.}
\end{deluxetable*}

\subsection{Flux Emergence Simulation (GT)} \label{subsec:gt}

As the GT data set, we used a self-consistent MHD simulation model of magnetic flux emergence. Typical flux emergence models simulate the process that an isolated magnetic flux tube, initially placed in the convection zone, buoyantly rises into the photosphere and eventually into the corona. The model we used in this study is based on the ``reference'' case of \citet{2017ApJ...850...39T}, where the generation mechanism of different types of flaring active regions observed by \citet{2017ApJ...834...56T} was theoretically investigated. We solved the full set of basic resistive MHD equations with the numerical code by \citet{2015ApJ...813..112T}, which calculates the spatial derivatives by the fourth-order central differences and the temporal derivatives by the four-step Runge-Kutta scheme based on \citet{2005A&A...429..335V}. This code also implements the divergence cleaning method by \citet{2002JCoPh.175..645D} to ensure $\nabla\cdot\mbox{\boldmath $B$}=0$. Physical values are normalized by $H_{0}=170\ {\rm km}$ for length, $C_{\rm s0}=6.8\ {\rm km\ s}^{-1}$ for velocity, $\tau_{0}=25\ {\rm s}$ for time, $B_{0}=250\ {\rm G}$ for magnetic field strength, etc.

The only differences from the previous case, namely, the ``reference'' case in \citet{2017ApJ...850...39T}, are the termination time of the computation and the box size. The reason of this change is that the evolution of the flux tube did not saturate at the termination time $t/\tau_{0}=300$ which we adopted before. Therefore, in the present study, we continued the calculation further until $t/\tau_{0}=500$ so that the total magnetic energy in the atmosphere and the total unsigned magnetic flux in the photosphere reach their saturation levels (see Sec. \ref{subsec:general}). In order to sufficiently cover the expanded flux tube at $t/\tau_{0}=500$, the box was expanded to $(-300, -300, -40)\leq (x/H_{0}, y/H_{0}, z/H_{0})\leq (300, 300, 500)$, namely, the box size was $L_{x}/H_{0}\times L_{y}/H_{0}\times L_{z}/H_{0}=600\times 600\times 540$, or $102\ {\rm Mm}\times 102\ {\rm Mm}\times 92\ {\rm Mm}$. The simulation box was resolved by the grid number of $N_{x}\times N_{y}\times N_{z}=720\times 720\times 650$ with the smallest grid spacing of $(\Delta x/H_{0}, \Delta y/H_{0}, \Delta z/H_{0})=(0.25, 0.25, 0.2)$ at the domain center, gradually increasing up to $(1.5, 1.5, 1.6)$.

The initial background atmosphere was gravitationally stratified and consisted of the adiabatically stratified convection zone ($z/H_{0}<0$), the cool isothermal photosphere/chromosphere ($0\leq z/H_{0}< 18$), and the hot isothermal corona ($z/H_{0}\geq 18$). The flux tube was originally embedded at $z=-30H_{0}=5,100\ {\rm km}$, directed along the $x$-axis with the form of $B_{x}(r)=B_{\rm tube}\exp{(-r^{2}/R_{\rm tube}^{2})}$ and $B_{\phi}(r)=qrB_{x}(r)$, where $r$ is the radial distance from the tube's axis, $R_{\rm tube}=3H_{0}=510\ {\rm km}$ the radius, $B_{\rm tube}=30B_{0}=7.5\ {\rm kG}$ the axial field strength, and $q=-0.2/H_{0}=-1.2\times 10^{-6}\ {\rm m}^{-1}$ the twist intensity (the negative sign indicates a left-handed twist). The total axial flux is $\Phi_{\rm tube}=850B_{0}H_{0}^{2}=6.1\times 10^{19}\ {\rm Mx}$. The middle of the tube around $x/H_{0}=0$ was made buoyant and the tube started emergence due to its own buoyancy.

As the time-dependent photospheric boundary data for the coronal field models, we extracted 2D slices at $z/H_{0}=0$ from the original GT simulation. We provided the slices of all components of magnetic and velocity vectors, i.e., $B_{x}$, $B_{y}$, $B_{z}$, $V_{x}$, $V_{y}$, and $V_{z}$. The slices were sampled at every $\Delta t/\tau_{0}=1$ from $t/\tau_{0}=0$ to 500. The slices spanned over $(-165, -165)\leq (x/H_{0}, y/H_{0})\leq (165, 165)$, resolved by the uniform grid spacing of $400\times 400$, were distributed to each of the reconstruction models as synthetic observables.\footnote{The GT dataset is available online at \url{https://doi.org/10.5281/zenodo.3591984}.}

\subsection{Coronal Field Models}

\subsubsection{E-driven Magnetofrictional Model (MF)}

The magnetofrictional method, first introduced by \citet{1986ApJ...309..383Y} and \citet{1986ApJ...311..451C}, assumes that the plasma velocity is proportional to the Lorentz force. In this study, we used the magnetofrictional code by \citet{2012ApJ...757..147C} and hereafter the result is referred to as MF. In the present model, the magnetic field is driven by the photospheric electric field $\mbox{\boldmath $E$}$, calculated from the sequential GT slices via Ohm's law, $\mbox{\boldmath $E$}=-\mbox{\boldmath $V$}\times \mbox{\boldmath $B$}/c$. However, as the property of the MHD code used in the GT simulation, the ideal $\mbox{\boldmath $E$}$ obtained directly from the GT data does not necessarily satisfy Faraday's law,
\begin{eqnarray}
  \frac{\partial \mbox{\boldmath $B$}}{\partial t}=-c\nabla\times\mbox{\boldmath $E$},
  \label{eq:faraday}
\end{eqnarray}
because the GT code is meant to evolve the magnetic field through the induction equation,
\begin{eqnarray}
  \frac{\partial \mbox{\boldmath $B$}}{\partial t}=\nabla\times(\mbox{\boldmath $V$}\times \mbox{\boldmath $B$}).
  \label{eq:induction}
\end{eqnarray}
Therefore, before implemented in the MF model, we corrected $\mbox{\boldmath $E$}$ by adding minor correction terms so that it satisfies Eq. (\ref{eq:faraday}).\footnote{More precisely, the GT code solves the quantities including \mbox{\boldmath $B$} and \mbox{\boldmath $V$} on the grid points (i.e., cell centers) and does not compute \mbox{\boldmath $E$} on the grids nor on the cell edges, as opposed to the MF model, which deals with $\mbox{\boldmath $E$}$ on cell edges and takes the curl of it to compute how $\mbox{\boldmath $B$}$ evolves on cell faces. In the GT code, the induction equation is implemented with the form of $\partial \mbox{\boldmath $B$}/\partial t=-\nabla\cdot (\mbox{\boldmath $V$}\mbox{\boldmath $B$}-\mbox{\boldmath $B$}\mbox{\boldmath $V$})$, and this does not require the exact inductivity. This is why $\mbox{\boldmath $E$}$ from the GT simulation, derived as $\mbox{\boldmath $E$}=-\mbox{\boldmath $V$}\times\mbox{\boldmath $B$}/c$, is not inductive and the correction is necessary. For the correction technique, see \citet{2012ApJ...757..147C}.}

The computation was performed in a cubic box of the size $L_{x}/H_{0}\times L_{y}/H_{0}\times L_{z}/H_{0}=330\times 330\times 330$, ranging over $(-165, -165, 0)\leq (x/H_{0}, y/H_{0}, z/H_{0})\leq (165, 165, 330)$, and the grid number was assigned as $400\times 400\times 400$ (uniform). The periodic and open boundary conditions are adopted to the side and top boundaries, respectively. The temporal integration started at $t/\tau_{0}=150$ and continued until 500.

\subsubsection{B-driven MHD Model (MHD1)}\label{sec:mhd1}

The first MHD reconstruction (hereafter MHD1) was done with the numerical code by \citet{2016NatCo...711522J,2016ApJ...828...62J}, which is based on the data-driven model by \citet{2006ApJ...652..800W}. This code solves the full set of the MHD equations by the CESE-MHD scheme \citep{2010SoPh..267..463J}, in which all MHD variables are specified at the corners of the grid cells and no ghost cells are used at the boundary. It directly takes in the photospheric $\mbox{\boldmath $B$}$ and $\mbox{\boldmath $V$}$ slices and advances the magnetic field forward in time by solving the induction equation (\ref{eq:induction}).

The initial conditions consisted of a zero magnetic field and a plasma in a hydrostatic state, stratified by the solar gravity with a uniform temperature $10^{6}\ {\rm K}$ (typical in the corona). Thus, it did not include the photosphere, the chromosphere, and the transition region. The bottom boundary of the model was assumed at the coronal base with a fixed density of $10^{9}\ {\rm cm}^{-3}$ (typical in the corona). As the input magnetic field data was assumed to be taken on the photosphere and no preprocessing was made, there was inconsistency between the input data and the model (previously such inconsistency was circumvented by data smoothing or preprocessing), which might be one cause of the drastic expansion of magnetic field (Sec. \ref{subsec:general}).

As the magnetic flux drastically expands in this case, the computation was performed in a much enlarged box. First, the input GT slices were rebinned to $200\times 200$ grids with a spacing of $1.65H_{0}$ or 280 km. Then, the data-driven model was performed in a 3D box with a block-based non-uniform grid. The whole domain has the size of $L_{x}/H_{0}\times L_{y}/H_{0}\times L_{z}/H_{0}=846\times 846\times 846$, spanning over $(-422, -422, 0)\leq (x/H_{0}, y/H_{0}, z/H_{0})\leq (422, 422, 845)$, with the smallest grid being $1.65H_{0}$.

On the bottom boundary, the magnetic field strength outside the input GT slice was assumed to be zero. At the side and top boundaries, the plasma density, pressure, and velocity were fixed to their original values. The tangential component of magnetic field was linearly extrapolated from the inner pixels, whereas the normal component was obtained from the solenoidal condition, $\nabla\cdot \mbox{\boldmath $B$}=0$. This choice of boundary condition allows the magnetic field to freely penetrate the boundaries.

As a preliminary trial with an attempt to reduce the drastic expansion, we also tested the MHD1 model with the GT slices at $z/H_{0}=10$ instead of $z/H_{0}=0$. The setup and results are shown in Appendix \ref{app:mhd1}.

\subsubsection{B-driven MHD Model (MHD2)}

For the second type of the MHD models, MHD2, we used the zero-$\beta$ MHD code introduced by \citet{2019ApJ...870L..21G}, which omits the gas pressure gradient and gravity and only solves the density, velocity, and magnetic field. This model is implemented in the Message-Passing Interface-Adaptive Mesh Refinement Versatile Advection Code (MPI-AMRVAC; \citealt{2012JCoPh.231..718K}). Similar to MHD1, the coronal field is driven by the photospheric magnetic field (i.e., magnetic field-driven model). In the present model, we left out the following three source terms: the artificial density diffusion in the mass conservation equation; the viscosity in the momentum conservation equation; and the resistivity in the induction equation. However, the divergence-cleaning source term was kept in the induction equation.

The initial background atmosphere was in a hydrostatic equilibrium with a cool photosphere/chromosphere of $10^{4}\ {\rm K}$ extending up to $\sim 2.5\ {\rm Mm}$ and a hot corona of $10^{6}\ {\rm K}$ above $\sim 10\ {\rm Mm}$, and the two atmospheric layers were connected by a transition region. The plasma (proton) number density drops from $1.4\times 10^{17}\ {\rm cm}^{-3}$ on the bottom boundary to $3.9\times 10^{11}\ {\rm cm}^{-3}$ on the top boundary.

The computation was done from $t/\tau_{0}=150$ to 500 in a box of $L_{x}/H_{0}\times L_{y}/H_{0}\times L_{z}/H_{0}=330\times 330\times 330$ that was resolved by a uniform $400\times 400\times 400$ grid. There were two ghost layers on each of all six boundaries. The boundary conditions are specified in the cell centers of the ghost layers. On the side and top boundaries, the boundary conditions for the density, velocity, and magnetic field were symmetric. Regarding the bottom boundary, the density was fixed at the original value for the two ghost layers, whereas the magnetic field and velocity followed the provided GT values for the inner ghost layer, and these values were provided by a zero gradient extrapolation for the outer layer. The normal component of the magnetic field for the outer ghost layer on the bottom and for both ghost layers on the side and top boundaries is modified to satisfy the divergence-free condition.

\subsubsection{E-driven MHD Model (MHD3)}

The third MHD model, MHD3, is based on \citet{2018ApJ...855...11H,2019ApJ...871L..28H}. In brief, this method calculates electric field vectors at three heights (center of bottom boundary cell, and the top and bottom interfaces), curl of which fully matches the temporal variations of three-component bottom boundary magnetic field given from the GT model. The plasma velocity is assumed to be zero and the plasma density and temperature are fixed on the bottom boundary surface. In this way, the data values in the ghost cells are not used for updating the simulated MHD variables on and above the bottom boundary at all. Driven with the electric field through Equation (\ref{eq:faraday}), the simulated magnetic field maintains the divergence-free condition in the simulation box all through the evolution.

Here we assumed a uniform, non-stratified initial background atmosphere with the constant plasma (proton) number density of $10^{13}\ {\rm cm}^{-3}$ and omitted the gravity.

Although the box size was consistent with most of the other cases (i.e., $L_{x}/H_{0}\times L_{y}/H_{0}\times L_{z}/H_{0}=330\times 330\times 330$), the grid number was reduced by a factor of 2 to $200\times 200\times 200$ (uniform) due to the limited computation speed of the code. In addition, extra buffer layers of 30-grid thickness were introduced in the horizontal directions in order to avoid computational difficulties at the edges on the bottom boundary surface, and hence the actual grid number of the simulated volume was $260\times 260\times 200$. On the bottom boundary, the simulated magnetic field vectors are updated sequentially in accordance with the input from the GT model field through the electric field vectors over the central $200\times 200$-grid area. The strength of the simulated magnetic field in the outer 30-grid thick layers is zero on the bottom boundary surface, but not for the rest of the height. The characteristics-based boundary conditions are applied to the top and side boundary surfaces to regulate the boundary treatment in a way similar to the open boundary condition for outflows. (Inflows are much less frequent on the side and top boundary surfaces in the simulated case.)

\section{Results} \label{sec:results}

In all cases of the coronal field reproduction, the outcome contains a twisted flux rope structure, consistent with the GT magnetic field, but the individual models exhibit variations to a certain degree. In this section, we show the temporal evolutions of basic parameters (Section \ref{subsec:general}), the overall coronal field structures (Section \ref{subsec:snapshot}), the evolutions of magnetic helicity (Section \ref{subsec:helicity}), and the degree of force-freeness (Section \ref{subsec:force-free}).

\subsection{General Evolution} \label{subsec:general}

\begin{figure*}
\begin{center}
\includegraphics[width=0.9\textwidth]{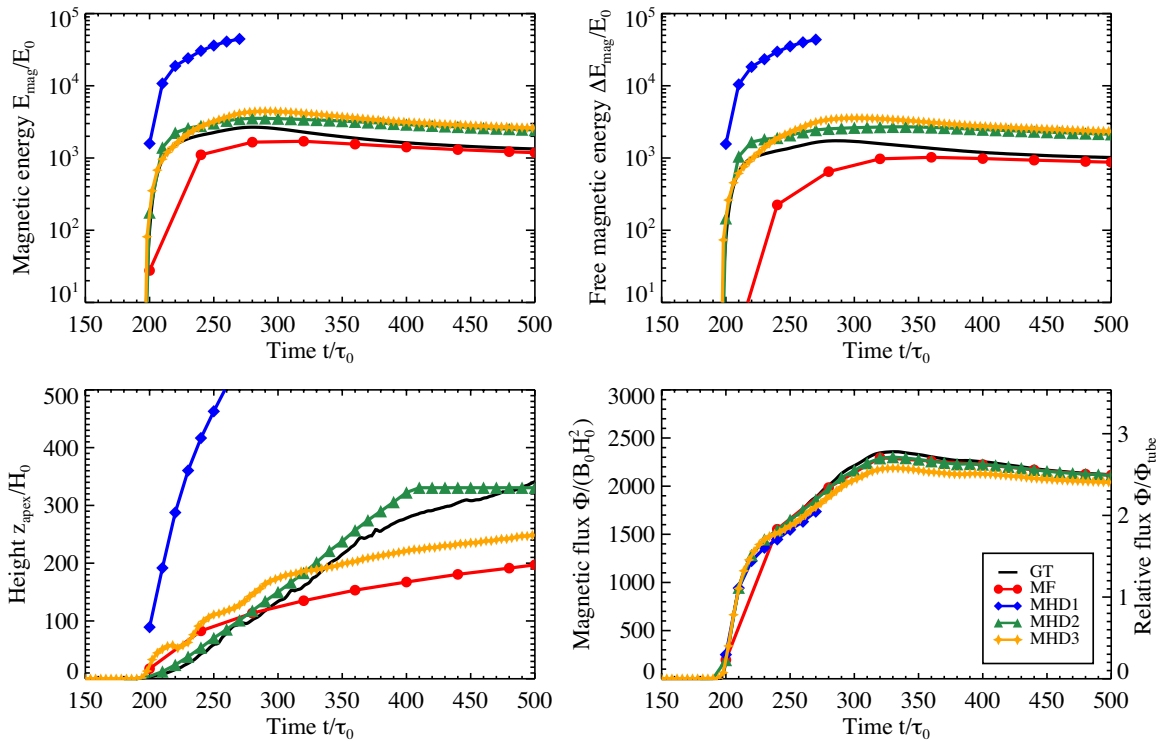}
\end{center}
\caption{Time evolutions of the magnetic energy $E_{\rm mag}/E_{0}$, free magnetic energy $\Delta E_{\rm mag}/E_{0}$, apex height $z_{\rm apex}/H_{0}$, and total unsigned photospheric flux $\Phi$ for GT and the four coronal field models, MF, MHD1, MHD2, and MHD3. In the bottom right panel, the right vertical axis presents the total unsigned flux normalized by the total axial flux of the initial flux tube $\Phi_{\rm tube}$.\label{fig:general}}
\end{figure*}

Figure \ref{fig:general} shows the temporal evolutions of the magnetic energy,
\begin{eqnarray}
  E_{\rm mag}=\int_{z\geq 0} \frac{\mbox{\boldmath $B$}^{2}}{8\pi}\, dV;
  \label{eq:emag}
\end{eqnarray}
the free magnetic energy,
\begin{eqnarray}
  \Delta E_{\rm mag} 
  = \int_{z\geq 0} \frac{\mbox{\boldmath $B$}^{2}}{8\pi}\, dV
  -\int_{z\geq 0} \frac{\mbox{\boldmath $B$}_{\rm PF}^{2}}{8\pi}\, dV,
  \label{eq:efre}
\end{eqnarray}
where $\mbox{\boldmath $B$}_{\rm PF}$ is the potential magnetic field; the apex height, defined in this study as the highest part of the emerging flux where the field strength exceeds the threshold value $|\mbox{\boldmath $B$}|/B_{0}=0.005$; and the total unsigned magnetic flux in the photosphere,
\begin{eqnarray}
  \Phi=\int_{z=0} |B_{z}|\, dS.
  \label{eq:phi}
\end{eqnarray}

In regard to the GT evolution, the magnetic energies, both $E_{\rm mag}$ and $\Delta E_{\rm mag}$, and the total unsigned flux reach their peak values around $t/\tau_{0}=280$ and 330, respectively. It is known that when the emerging magnetic fields appear in the photosphere, they tend to take an undular shape, wandering up and down across the surface to increase the unsigned flux, even devoid of the thermal convection \citep[e.g.,][]{2007ApJ...657L..53I,2009A&A...508.1469A}. This may be the reason why the photospheric unsigned flux is greater than twice of the tube's original axial flux ($\Phi/\Phi_{\rm tube}\sim 2.7$). As time goes on, however, these parameters turn into a gradual reduction phase due to a further free expansion into the corona. Although the apex height of GT shows a monotonic increase over the whole time period, the rise velocity decreases around $t/\tau_{0}=370$ after the photospheric flux levels off. The final height at $t/\tau_{0}=500$ is $z_{\rm apex}/H_{0}=342$.

The temporal evolutions of the magnetic energies ($E_{\rm mag}$ and $\Delta E_{\rm mag}$) for the coronal field reconstruction models are, except for MHD1, in good agreement with the GT trends. In the later declining phase ($t/\tau_{0}\gtrsim 350$), the energies of MF converge to those of GT, whereas those of MHD2 and MHD3 are larger than GT by a factor of two.

The apex height ($z_{\rm apex}/H_{0}$) also shows some model dependence. MHD2 keeps up with the GT curve until GT saturates around $t/\tau_{0}=370$ and, because the flux rope hits the top boundary, MHD2 levels off at $z_{\rm apex}/H_{0}=330$. Although the initial speeds of MF and MHD3 are faster than that of GT, they quickly slow down and end up with $z_{\rm apex}/H_{0}=197$ and 249, respectively.

One can see from these diagrams that the most drastic evolution appears for MHD1. In this case, both $E_{\rm mag}$ and $\Delta E_{\rm mag}$ are larger than those for GT by more than an order of magnitude. The summit reaches $z_{\rm apex}/H_{0}=360$ already at $t/\tau_{0}=230$ and eventually exceeds 540 at $t/\tau_{0}=270$.

In theory, the photospheric unsigned fluxes ($\Phi$) for the reconstruction models should be identical to that of GT, and the bottom right panel of Figure \ref{fig:general} demonstrates that this is indeed true for most of the times.

\subsection{Overall Field Line Structures}\label{subsec:snapshot}

\begin{figure*}
\begin{center}
\includegraphics[width=0.85\textwidth]{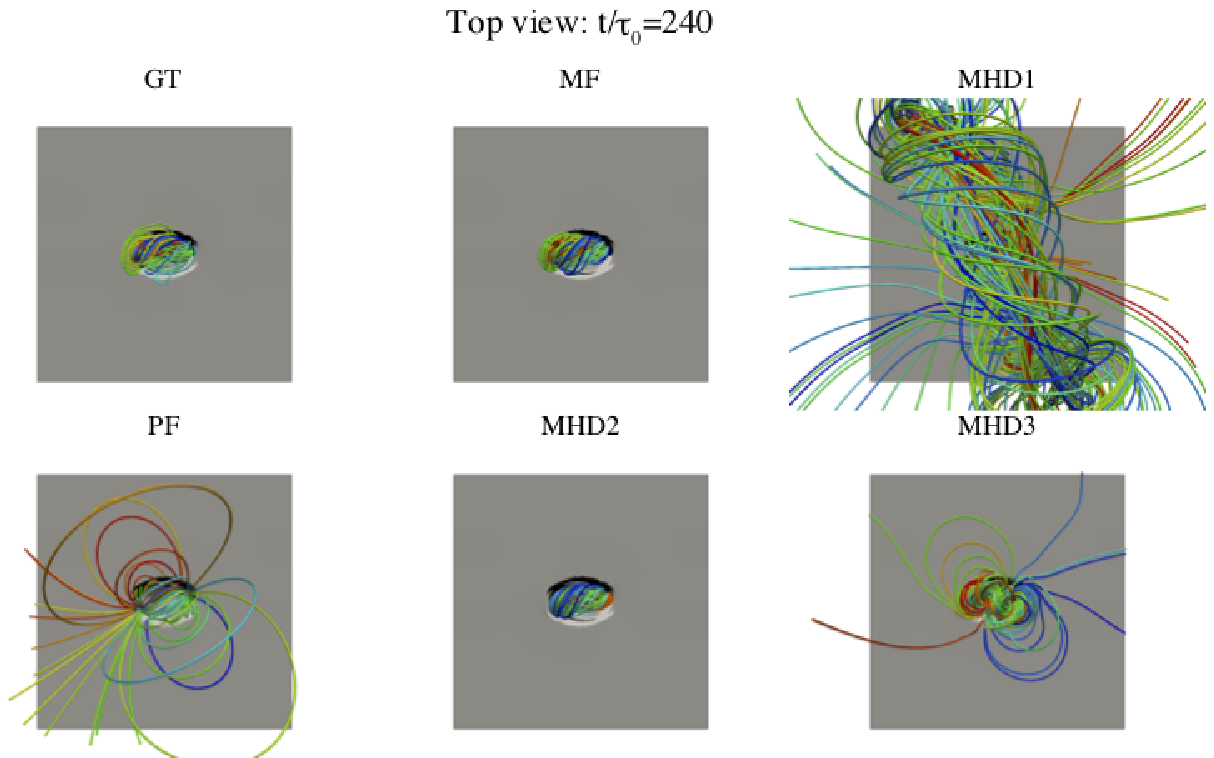}
\end{center}
\caption{Top view of magnetic fields for GT, PF, and the four coronal field reconstruction models at $t/\tau_{0}=240$. The bottom boundary shows the vertical GT magnetic field (i.e., $B_{z}/B_{0}$ at $z/H_{0}=0$), saturating at $-0.1$ (black) and $0.1$ (white). The tubes indicate the field lines, where the tubes with reddish (bluish) colors are integrated from the seeds placed in the positive (negative) polarity. The seeds are identical for all six cases.\label{fig:0240top}}
\end{figure*}

\begin{figure*}
\begin{center}
\includegraphics[width=0.85\textwidth]{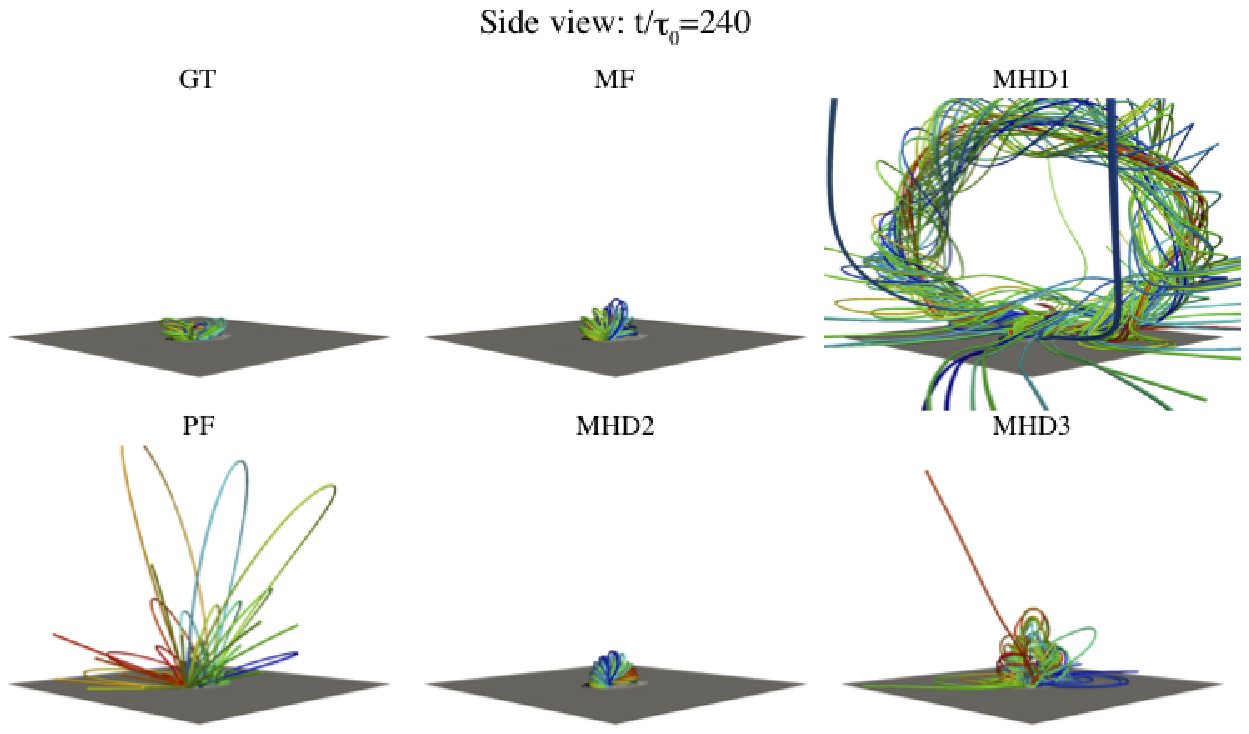}
\end{center}
\caption{The same as Figure \ref{fig:0240top} but for the side views. An animated version of this figure is available, which presents the side views from different angles.\label{fig:0240side}}
\end{figure*}

\begin{figure*}
\begin{center}
\includegraphics[width=0.85\textwidth]{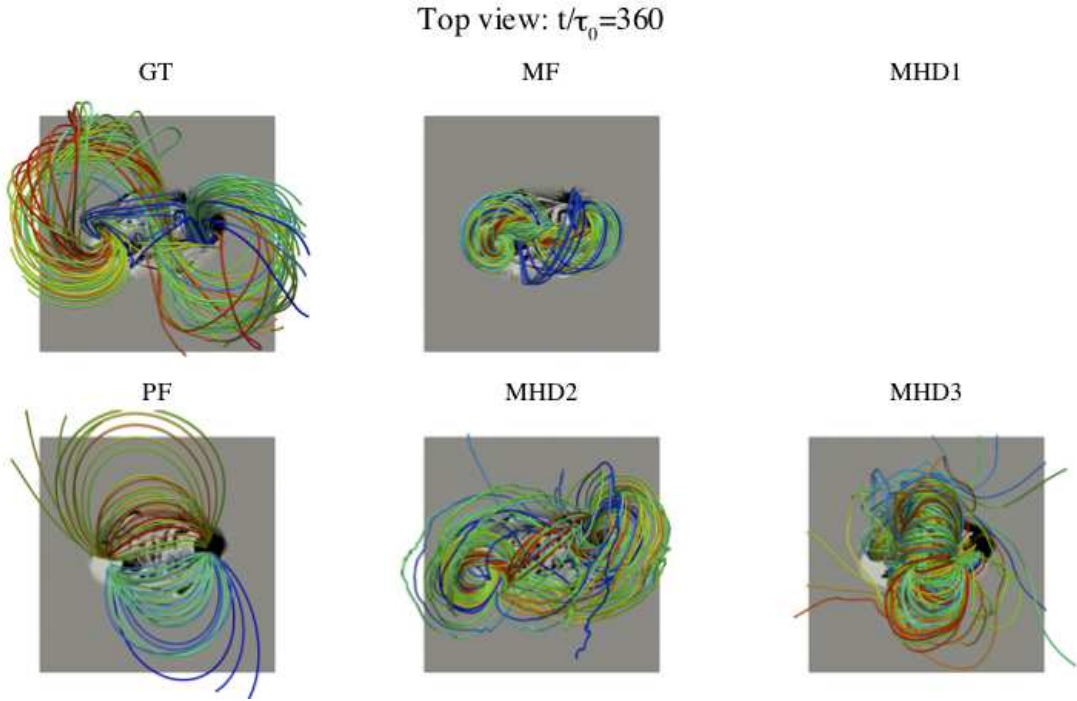}
\end{center}
\caption{The same as Figure \ref{fig:0240top} but for $t/\tau_{0}=360$. MHD1 is not shown because the computation is terminated before this moment.\label{fig:0360top}}
\end{figure*}

\begin{figure*}
\begin{center}
\includegraphics[width=0.85\textwidth]{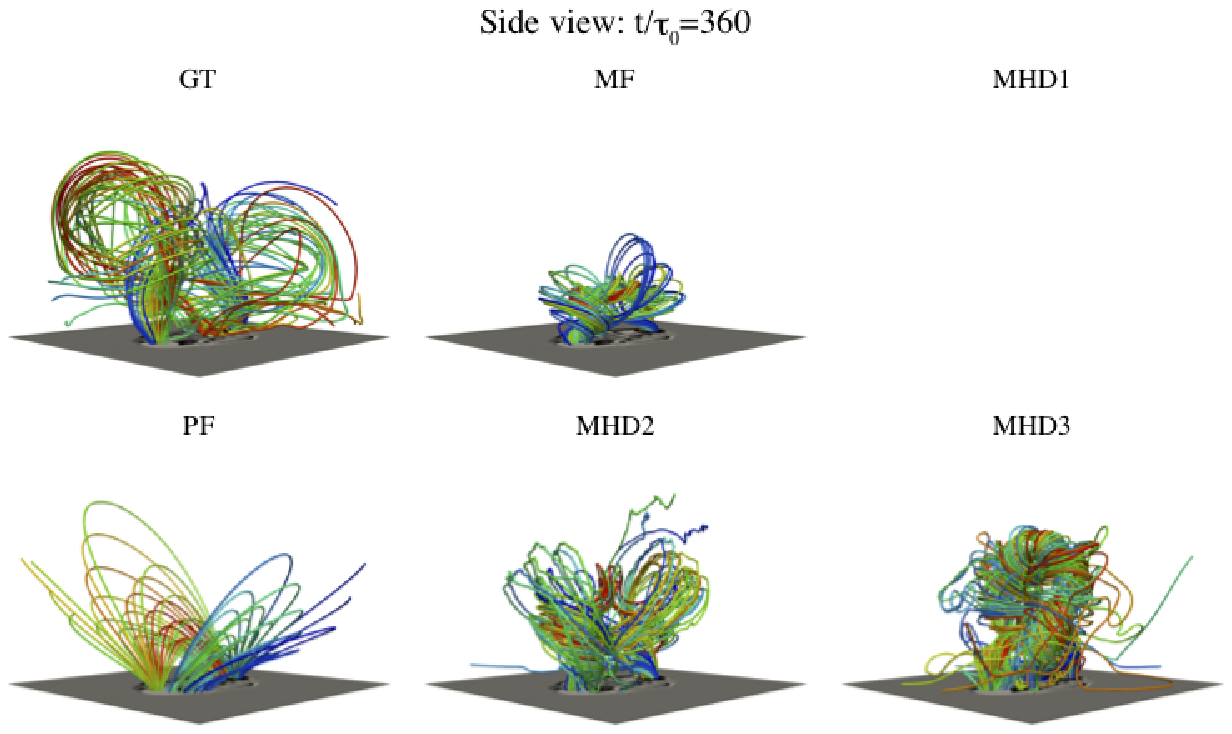}
\end{center}
\caption{The same as Figure \ref{fig:0360top} but for the side views. An animated version of this figure is available, which presents the side views from different angles.\label{fig:0360side}}
\end{figure*}

Figures \ref{fig:0240top}--\ref{fig:0360side} are the top and side views of GT, PF, and the four coronal field models for $t/\tau_{0}=240$ and 360. The PF is extrapolated from the vertical component of the GT magnetic slices on the photosphere. Magnetic field lines, as represented by colored tubes, are integrated from the seeds in the photosphere (reddish for positive polarity and bluish for negative), and for each time step, the field lines with the same color share a common photospheric seed.

At $t/\tau_{0}=240$, when the apex of the emerging flux steps into the coronal volume (see the top left panels of Figures \ref{fig:0240top} and \ref{fig:0240side}), the top and side images of GT, MF, and MHD2 present a low-lying, slightly helical magnetic dome that connects the two flux concentrations in the photosphere. The structures look similar to each other especially when seen from above (see the top views). The main body of the magnetic field structure of MHD3 appears to be consistent with GT, but it is more complicated with the laterally extended wings beside. Already by this moment, MHD1 has established a highly twisted magnetic torus that stands vertically in the atmosphere. Whereas the GT magnetic flux remains at the height of $z_{\rm apex}/H_{0}=43$ at this time, this remarkable feature reaches 416, i.e., about 10 times higher than GT.

At $t/\tau_{0}=360$, GT shows a well-developed flux rope connecting the two photospheric footpoints with an inverse S-shaped sigmoid (see especially the top view of Figure \ref{fig:0360top}). The inverse S is a natural consequence of the emergence of a left-handed flux tube. It is worth noting here that the flux rope expands not only in the vertical direction but also, or even more, in the lateral directions. This horizontal expansion occurs because the plasma drains down along the field lines due to the gravity and is typical of MHD flux emergence simulations \citep[see, e.g.,][]{1989ApJ...345..584S,2011ApJ...735..126T}. The inverse S-shaped sigmoids are clearly reproduced in the MF and MHD2 models. As the side views reveal, the flux rope remains lower down for the MF model (see also bottom left panel of Figure \ref{fig:general}), while the upper parts of MHD2 possess jagged field lines probably due to numerical errors. In spite of the coarse grid spacing, MHD3 manages to reproduce a helical flux rope, although the horizontal expansion is not prominent and thus the sigmoidal shape is not quite obvious.

\subsection{Magnetic Helicity} \label{subsec:helicity}

To assess if the transport of magnetic twist from (below) the photosphere to the corona is accurately reproduced, we measure the relative magnetic helicity for all simulation cases. The relative helicity of magnetic field $\mbox{\boldmath $B$}$ with respect to its reference potential magnetic field $\mbox{\boldmath $B$}_{\rm PF}$ is given by
\begin{eqnarray}
  H_{\rm R} = \int_{V} (\mbox{\boldmath $A$}+\mbox{\boldmath $A$}_{\rm PF})
  \cdot(\mbox{\boldmath $B$}-\mbox{\boldmath $B$}_{\rm PF})\, dV,
  \label{eq:helicity}
\end{eqnarray}
where $\mbox{\boldmath $A$}$ is the vector potential of $\mbox{\boldmath $B$}$ (i.e., $\mbox{\boldmath $B$}=\nabla\times\mbox{\boldmath $A$}$) and $\mbox{\boldmath $A$}_{\rm PF}$ is that of $\mbox{\boldmath $B$}_{\rm PF}$.

Here we follow the computation method by \citet{2012SoPh..278..347V} for a Cartesian domain $V=[x_{1}, x_{2}]\times[y_{1}, y_{2}]\times[z_{1}, z_{2}]$, and the notations below are based on \citet{2015A&A...582A..76S}. Because $H_{\rm R}$ is gauge invariant, we are free to choose the gauge $\mbox{\boldmath $A$}\cdot\hat{\mbox{\boldmath $z$}}=0$ such that it satisfies
\begin{eqnarray}
  \mbox{\boldmath $A$}=\mbox{\boldmath $A$}_{0}
  -\hat{\mbox{\boldmath $z$}}\times
  \int_{z_{1}}^{z}\mbox{\boldmath $B$}(x, y, z')\, dz',
\end{eqnarray}
where $\mbox{\boldmath $A$}_{0}=\mbox{\boldmath $A$}(x, y, z=z_{1})=(A_{0x}, A_{0y}, 0)$ is a solution to the $z$-component of $\mbox{\boldmath $B$}=\nabla\times\mbox{\boldmath $A$}$. We take one simple solution to this equation:
\begin{eqnarray}
  A_{0x}=-\frac{1}{2} \int_{y_{1}}^{y} B_{z}(x, y', z=z_{1})\, dy',\nonumber\\
  A_{0y}=\frac{1}{2} \int_{x_{1}}^{x} B_{z}(x', y, z=z_{1})\, dx'.\nonumber
\end{eqnarray}
The vector potential for $\mbox{\boldmath $B$}_{\rm PF}$ is similarly calculated using the common $\mbox{\boldmath $A$}_{0}$ as
\begin{eqnarray}
  \mbox{\boldmath $A$}_{\rm PF}=\mbox{\boldmath $A$}_{0}
  -\hat{\mbox{\boldmath $z$}}\times
  \int_{z_{1}}^{z}\mbox{\boldmath $B$}_{\rm PF}(x, y, z')\, dz'.
\end{eqnarray}

\begin{figure}
\begin{center}
\includegraphics[width=90mm]{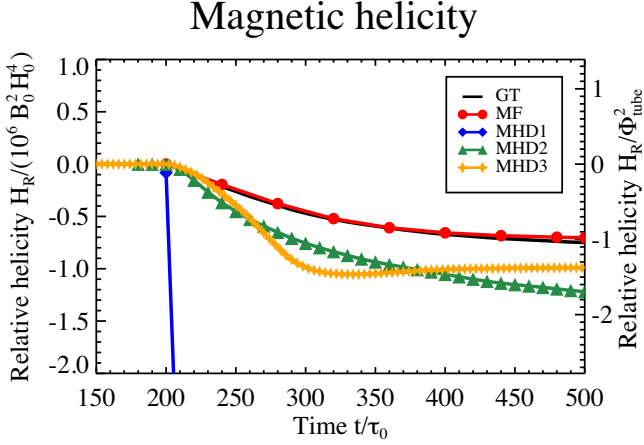}
\end{center}
\caption{Time evolutions of the relative magnetic helicity $H_{\rm R}$ for GT and the four coronal field models, MF, MHD1, MHD2, and MHD3. The right vertical axis presents the relative helicity normalized by the square of the total flux of the initial flux tube $\Phi_{\rm tube}^{2}$.\label{fig:helicity}}
\end{figure}

Figure \ref{fig:helicity} compares the evolutions of the relative magnetic helicity, $H_{\rm R}$, measured by the above method. The computation domain is $(-423, -423, 0)\leq (x/H_{0}, y/H_{0}, z/H_{0})\leq (423, 423, 846)$ for MHD1 and $(-165, -165, 0)\leq (x/H_{0}, y/H_{0}, z/H_{0})\leq (165, 165, 330)$ for all the other cases. One can find from this figure that $H_{\rm R}$ for GT increases monotonically on the negative side, and this is natural as we assumed the original flux tube with a negative (i.e. left-handed) twist (see Sec. \ref{subsec:gt}). The vertical axis on the right shows the relative helicity normalized by the square of the initial total axial flux, $H_{\rm R}/\Phi_{\rm tube}^{2}$. The value for GT reaches $\sim -1$, which corresponds to that the field lines of almost one full turn of the original flux tube are emerged in the atmosphere.

The GT curve is almost perfectly reproduced by MF, with the maximum deviation being only 10\%. The helicity of MHD2 is up to 1.7 times of GT. Together with the fact that the magnetic energies of MHD2, both $E_{\rm mag}$ and $\Delta E_{\rm mag}$, are about twice the GT values (see top panels of Figure \ref{fig:general}), this may imply an extra helicity injection continually throughout the evolution. The MHD3 trend also shows an overestimation but with a deflection around $t/\tau_{0}=300$ and the eventual recovery. Regarding MHD1, the helicity already amounts to $H_{\rm R}/(B_{0}^{2}H_{0}^{4})=-3.6\times 10^{6}$ at $t/\tau_{0}=210$ and, at $t/\tau_{0}=270$, it reaches $42.1\times 10^{6}$, about 120 times the GT value.

\subsection{Force-freeness} \label{subsec:force-free}

\begin{figure}
\begin{center}
\includegraphics[width=90mm]{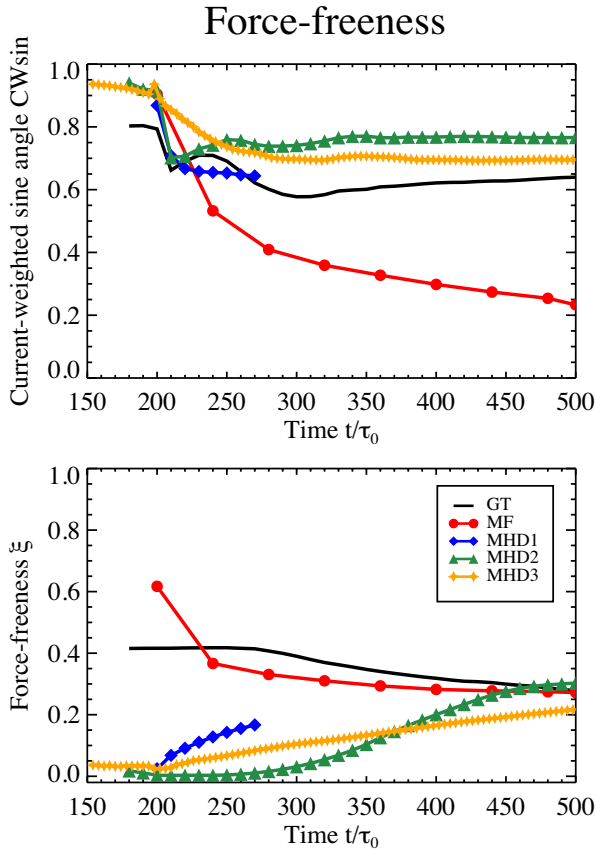}
\end{center}
\caption{Time evolutions of the domain-averaged current-weighted sine of the angle between $\mbox{\boldmath $B$}$ and $\mbox{\boldmath $j$}$, ${\rm CWsin}$, and the metric $\xi$ for GT and the four coronal field models, MF, MHD1, MHD2, and MHD3.\label{fig:force-free}}
\end{figure}

We also investigate how force free the GT and the reconstructed magnetic fields are by measuring the two following metrics. The first parameter is the domain-averaged current-weighted sine of the angle between the magnetic field $\mbox{\boldmath $B$}$ and the electric current $\mbox{\boldmath $j$}$,
\begin{eqnarray}
  {\rm CWsin}=
  \frac{\sum |\sin{\mu}| |\mbox{\boldmath $j$}|}{\sum |\mbox{\boldmath $j$}|},
\end{eqnarray}
where $\mu$ is the angle between $\mbox{\boldmath $B$}$ and $\mbox{\boldmath $j$}$. The magnetic field is force free when ${\rm CWsin} \ll 1$, but the opposite is not necessarily true for the increased ${\rm CWsin}$.

In fact, the temporal evolutions of this parameter shown in the top panel of Figure \ref{fig:force-free} do not present a clear trend. Whereas ${\rm CWsin}$ for MF decreases monotonically and drops from 0.9 to 0.2, the other curves do not exhibit well defined trends, with some showing oscillations. The problem of this metric is the effect of the current-free subregions within the computational domain, where $\mbox{\boldmath $j$}=(c/4\pi)\nabla\times\mbox{\boldmath $B$}$ can be non-zero due to numerical error.

To overcome this issue, \citet{2014ApJ...783..102M} proposed an alternative metric for the force-freeness that is not sensitive to the absence of currents:
\begin{eqnarray}
  \xi=\frac{1}{N} \sum_{i=1}^{N}
  \frac{|\mbox{\boldmath $F$}_{\rm L}|}
       {|\mbox{\boldmath $F$}_{\rm mp}|+|\mbox{\boldmath $F$}_{\rm mt}|},
\end{eqnarray}
where $\mbox{\boldmath $F$}_{\rm L}$, $\mbox{\boldmath $F$}_{\rm mp}$, and $\mbox{\boldmath $F$}_{\rm mt}$ are the Lorentz force, magnetic pressure gradient, and magnetic tension, respectively, and $N (=N_{x}\times N_{y}\times N_{z})$ stands for the total grid number. In other words, the metric $\xi$ represents the domain average of the Lorentz force relative to its components, and the field is force free when $\xi \ll 1$, while the substantial Lorentz force resides in the volume when $\xi\sim 1$.

The bottom panel of Figure \ref{fig:force-free} presents a clear trend that for all simulation cases, the metric $\xi$ converges to $\sim 0.3$. The decreasing trend of GT and MF is due to the free expansion of the emerging flux into the corona, becoming more and more force-free. The remaining three curves, i.e., MHD1, MHD2, and MHD3, start from $\sim 0$, indicating that the model fields are almost perfectly force-free, however as the field is advected more into the domain, the values approach to the GT value of $\sim 0.3$.

\section{Summary and Discussion} \label{sec:discussion}

In this study, we have investigated the different types of data-driven coronal field models by leveraging an MHD flux emergence simulation as a reference, aiming at demonstrating the characteristics of the models. As a result of the qualitative and quantitative assessment, it was revealed that, at least, a helical flux rope is reproduced in all coronal field models examined. The key findings include:
\begin{itemize}
\item For MF, MHD2, and MHD3, the magnetic energies and relative magnetic helicity of the coronal field models are comparable to or at most twice as much as the GT values. The MHD models consistently overestimate both the energy and helicity of the GT field for reasons that are not clear.
\item For MF, MHD2, and MHD3, the apex height of the developing flux rope varies from about the same to half of the GT value.
\item The GT's sigmoidal flux rope structure is well reproduced by MF and MHD2, although a variation in size and jaggy field lines due to numerical errors are found for these cases. MHD3 shows a twisted flux rope, too, but devoid of a sigmoidal structure because the lateral expansion is not significant.
\item In the case of MHD1, a vertically standing magnetic torus with a large degree of twist is rapidly created.
\end{itemize}

Overall, the data-driven models exhibit a certain level of qualitative agreement, e.g., the reproduction of a highly twisted flux rope in the atmosphere. In aspects of quantitative evaluation, however, the physical parameters do not necessarily converge to the GT values. In what follows, we discuss the possible causes of the discrepancy between the GT and coronal field models, ascribing them to the input boundary condition and model constraints.

The first issue we address is that the GT simulation is highly dynamical and the photospheric magnetic field, i.e., the input bottom boundary to the coronal field models, is largely deviated from the force-free state. Although many of the coronal field models including extrapolation, data-constrained, and data-driven algorithms assume the force-freeness of the magnetic field (see Section \ref{sec:intro}), the reality is not the case because the photosphere is the realm where the gas pressure gradient and gravity exert an overwhelming influence. \citet{1985svmf.nasa...49L} pointed out that for a magnetic field in a half volume ($z\geq 0$) that quickly decays with height, the field is force-free when all three components of the net Lorentz force is much smaller than the volume integrated Lorentz force, $F_{x}\ll F_{0}$, $F_{y}\ll F_{0}$, and $F_{z}\ll F_{0}$, where
\begin{eqnarray}
  F_{x}=-\frac{1}{4\pi} \int B_{x}\, B_{z}\, dS,\nonumber
\end{eqnarray}
\begin{eqnarray}
  F_{y}=-\frac{1}{4\pi} \int B_{y}\, B_{z}\, dS,\nonumber
\end{eqnarray}
\begin{eqnarray}
  F_{z}=-\frac{1}{8\pi} \int \left( B_{z}^{2}-B_{x}^{2}-B_{y}^{2} \right) dS,\nonumber
\end{eqnarray}
and
\begin{eqnarray}
  F_{0}=\frac{1}{8\pi} \int \left( B_{x}^{2}+B_{y}^{2}+B_{z}^{2} \right) dS.\nonumber
\end{eqnarray}

\begin{figure}
\begin{center}
\includegraphics[width=90mm]{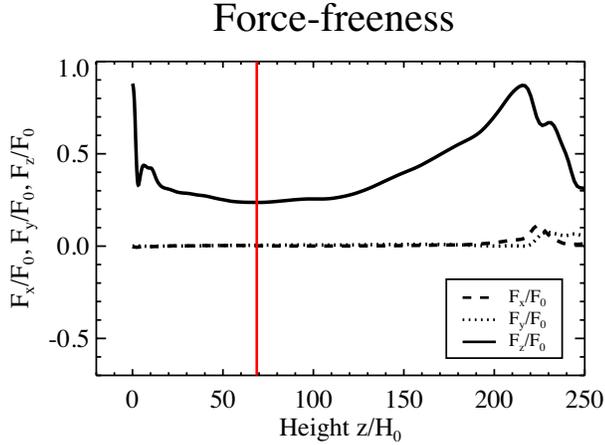}
\end{center}
\caption{Scaled net Lorentz force for the $x$ (dashed), $y$ (dotted), and $z$ (solid) components, measured for the GT magnetic field at $t/\tau_{0}=360$. Smaller magnitudes indicate that the field is more force-free, while the vertical red line shows the height at which $|F_{z}/F_{0}|$ takes its minimum.\label{fig:forcefreeness}}
\end{figure}

Figure \ref{fig:forcefreeness} demonstrates the height variations of $F_{x}/F_{0}$, $F_{y}/F_{0}$, and $F_{z}/F_{0}$. It was suggested by \citet{1995ApJ...439..474M} that the field is considered to be force-free if $|F_{z}/F_{0}|<0.1$. As one can see from this figure, the net Lorentz force measured at $z/H_{0}=0$ implies a great departure from the force-free state, reflecting the fact that, in the photosphere, the magnetic field evolves by the Lorentz force pushing against the pressure gradient and gravity. A bit higher up, however, the $F_{z}/F_{0}$ curve makes a dramatic drop to a much more force-free state, eventually reaching $F_{z}/F_{0}=0.24$ at $z/H_{0}=68.7$. Therefore, to surmount the difficulty of the non-force-free photosphere, one may introduce the magnetic information of, say, the upper photosphere or the chromosphere \citep{2019ApJ...870..101F}. Our preliminary MHD1 modeling based on the slices at $z/H_{0}=10$ instead of $z/H_{0}=0$ in Appendix \ref{app:mhd1} clearly yields a much more compact magnetic dome that is closer to the GT result. The application of non-force-free models, which allows the background atmosphere to be stratified or dynamically evolving, may be another choice, as we observed that MHD2 and MHD3, equipped with stratified and denser atmospheres, respectively, suppress the strong and fast flux injection.

The second point to be noted is that numerical conditions assumed in the coronal field models may inhibit the successful reproduction. For instance, in several models, we observed the overaccumulation of magnetic energy and helicity and the flux rope structure having a greater amount of magnetic twist. Such an excessiveness may derive from the model constraint that the photospheric field is fixed constant (or interpolated) during a time integration between the sequential updates of the photospheric boundary. The extra helicity can be deposited to the computational domain because the coronal field is rearranged during the time integration in such a way that the force-free $\alpha$ (Eq. (\ref{eq:alpha})), determined by the fixed footpoint vector field, becomes constant along each field line. Consequently, the coronal field can be overly twisted up. Examples of such situations are the NLFFF model in Appendix \ref{app:nlfff} and the data-constrained model of \citet[see their Figure 5]{2018ApJ...866...96J}. For data-driven models, it is therefore necessary to provide the appropriate treatment of the bottom boundary that allows the boundary field to be freely reconfigured, even between the updates, in response to the coronal field evolution. (For instance, the bottom field becomes more vertical, not fixed, when the coronal field expands upward.)

\begin{figure}
\begin{center}
\includegraphics[width=90mm]{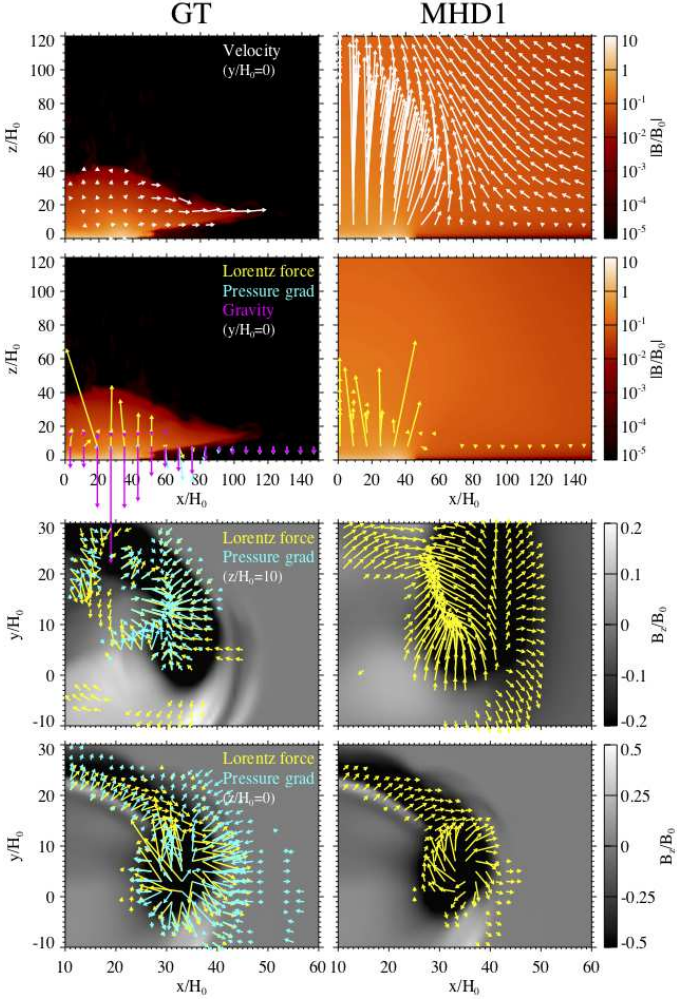}
\end{center}
\caption{Comparison of GT and MHD1 at $t/\tau_{0}=240$. (Top) Plasma velocity on the vertical plane $y/H_{0}=0$ plotted on the $|\mbox{\boldmath $B$}|$ field. (Second) Lorentz force, pressure gradient, and gravity on the same plane. (Third and bottom) Lorentz force and pressure gradient at the heights of $z/H_{0}=10$ and 0, respectively, plotted on the $B_{z}$ field. The strong upflow is excited in MHD1 because the Lorentz force around the bottom boundary is not counteracted by the pressure gradient and gravity.\label{fig:lorentz}}
\end{figure}

Regarding MHD1, the velocity field presents a significant difference between the bottom boundary and upper layers. As shown in the top panels of Figure \ref{fig:lorentz}, whereas the velocity at the bottom of MHD1 is consistent with that of GT with the typical values of $\lesssim 1C_{\rm s0}$ (or $6.8\ {\rm km\ s}^{-1}$), there exists a strong upflow of up to more than $50C_{\rm s0}$ ($340\ {\rm km\ s}^{-1}$) in the atmosphere that is not seen in GT. This discrepancy arises from the strong Lorentz force around the bottom boundary. In the case of GT, the Lorentz force exhibits a rotational, diverging, upward pattern, and this force is counteracted by the pressure gradient and gravity (see second to bottom panels). However, in MHD1, because the plasma is too weak to balance the Lorentz force, the flux rope is freely twisted and a strong upward motion is excited by the Lorentz force, which leads to the strong magnetic energy and helicity in the atmosphere.

Another constraint in the modeling is the spatial resolution. Interestingly, MHD3 managed to reproduce the magnetic flux rope in the atmosphere. Perhaps this is because the unresolved small-scale features may not affect the large-scale coronal topology. But still, the constructed flux rope did not possess a sigmoidal shape but instead yield a packed morphology, and the measured physical quantities were to some degree, though not largely, deviated from the GT values, indicating that the resolution does affect the computational results. In fact, \citet{2015ApJ...811..107D} revealed that the NLFFF results depend highly on the spatial resolution: the free energy becomes larger with increasing resolution, whereas the relative helicity values vary significantly between resolutions \citep[see also][]{2016SSRv..201..147V}.

From the viewpoint of applying data-driven models to actual observational data, the lower temporal cadence of the photospheric boundary data may cause additional issue. \citet{2017ApJ...838..113L} pointed out that for rapidly evolving features such as emerging flux, undersampling of the dynamics generates large electric currents and incorrect coronal fields and energies. Moreover, we should be aware that the observational data inherently contain some uncertainties (e.g., noise and 180$^{\circ}$ ambiguity). In this study, we attempted to understand the characteristics of the models, and the evaluation of such resolution dependencies and uncertainty effects requires further investigations that we defer to later publications.

In the near future, facilitated by the advancement of observational techniques, it is expected that the magnetic measurements in the upper atmospheric layers will be improved further. Such opportunities may provide the means to solve the issues we confronted in the present study, which eventually leads to a better understanding of magnetic structure and its evolution.


\acknowledgments

The authors wish to thank the anonymous referee for encouraging and insightful comments.
We acknowledge the support from Nagoya University for ISEE/CICR Workshop on Data-Driven Models of the Solar Progenitors of Space Weather and Space Climate (PIs: M.C.M. Cheung and K. Kusano).
This work was supported by JSPS KAKENHI Grant Numbers JP15H05814 (PI: K. Ichimoto), 16J02063 and 18K13579 (PI: S. Takasao), and by the NINS program for cross-disciplinary study (Grant Numbers 01321802 and 01311904) on Turbulence, Transport, and Heating Dynamics in Laboratory and Astrophysical Plasmas: ``SoLaBo-X''.
Numerical computations were in part carried out on Cray XC50 at Center for Computational Astrophysics, National Astronomical Observatory of Japan.
M.C.M.C. acknowledges funding from NASA's Heliophysics Grand Challenges Research grant ``Physics and Diagnostics of the Drivers of Solar Eruptions'' (NNX14AI14G to the LMSAL) and computing resources from NASA High-End Computing programme through the NASA Advanced Supercomputing Division at Ames Research Center.
C.J. acknowledges the support by NSFC (41822404, 41731067, 41574170) and his computational work was carried out on TianHe-1 (A) at the National Supercomputer Center in Tianjin, China.
Y.G. is supported by NSFC (11773016, 11733003, 11533005, and 11961131002) and his numerical calculations are conducted with the computing facilities in the High Performance Computing Center (HPCC), Nanjing University.

%

\vspace{5mm}





\appendix

\section{NLFFF Extrapolation}\label{app:nlfff}

The NLFFF extrapolation was computed by the MHD relaxation technique by \citet{2014ApJ...788..182I}. For each time step, we first calculated the initial-guess coronal field by a PF extrapolation from the vertical component of the GT magnetic field slices (i.e. $B_{z}$ at $z/H_{0}=0$). Then, the horizontal components ($B_{x}$ and $B_{y}$) were gradually added to the bottom boundary until all three components matched the input GT data. By directly solving the zero-$\beta$ MHD equations, the coronal field was calculated so that it satisfied the NLFFF condition, namely, the force-free $\alpha$ (Eq. (\ref{eq:alpha})) is invariant along each field line.

The domain has the size of $L_{x}/H_{0}\times L_{y}/H_{0}\times L_{z}/H_{0}=330\times 330\times 330$ and the gird number of $N_{x}\times N_{y}\times N_{z}=400\times 400\times 400$ (uniform). The boundary condition for the magnetic field on the side and top boundary was that the field strength was fixed at the initial guess, i.e., the potential magnetic field.

\begin{figure}
\begin{center}
\includegraphics[width=80mm]{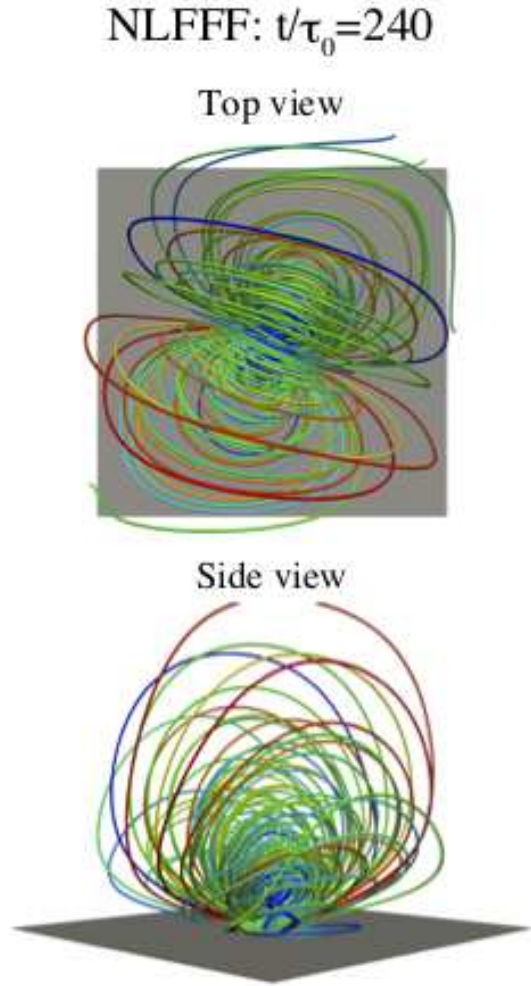}
\end{center}
\caption{Magnetic field at $t/\tau_{0}=240$ calculated by the NLFFF extrapolation technique. Like the other plots in Section \ref{sec:results}, the photospheric field ($B_{z}/B_{0}$ at $z/H_{0}=0$) is shown as a grayscale and the field lines are represented by colored tubes.\label{fig:nlfff}}
\end{figure}

Figure \ref{fig:nlfff} shows the snapshot at $t/\tau_{0}=240$. Although the GT flux rope has just entered the corona at this time, the NLFFF results present a well developed, highly twisted structure that fills the computational domain. The measured parameters for the NLFFF (GT) model at $t/\tau_{0}=240$ are: $E_{\rm mag}/E_{0}=6818$ (2072); $\Delta E_{\rm mag}/E_{0}=5935$ (1260); $z_{\rm apex}/H_{0}=210$ (43); $\Phi/(B_{0}H_{0}^{2})=1545$ (1542); $H_{\rm R}/(B_{0}^{2}H_{0}^{4})=-8.2\times 10^{5}$ ($-2.1\times 10^{5}$); ${\rm CWsin}=0.78$ (0.71); and $\xi=0.15$ (0.42). All these results indicate a strong accumulation of magnetic energy and twist like MHD1, albeit not that strong.

Although the iteration of NLFFF starts from PF, the resultant parameters largely surpass those of GT. This may be because, as a virtue of the force-free model, the magnetic twist along each field line tends to be uniform and follow the force-free $\alpha$ that is determined at the bottom boundary, whereas in the GT model it may not be true because of the strong stratification. This may lead to the overestimation of helicity and energy in the coronal volume.

For checking if the reproduced field is relaxed enough, we tracked the volume integral of Lorentz force (i.e. $\int |\mbox{\boldmath $j$}\times\mbox{\boldmath $B$}|\, dV$) over the iteration. However, the present result did not reach the saturation level within a trackable finite time, which indicates that the photospheric surface does not provide a suitable input for the force-free extrapolations.

\section{MHD1 Based on the $z/H_{0}=10$ Slices}\label{app:mhd1}

For further comparison, we also reproduced the coronal field with the MHD1 code but based on the $\mbox{\boldmath $B$}$ and $\mbox{\boldmath $V$}$ slices at $z/H_{0}=10$. The computation was done with exactly the same grid settings, initial and boundary conditions as described in Section \ref{sec:mhd1}. To suppress the expansion, the field strength was reduced to 20\% of the original value, and the bottom boundary was still assumed at the base of the corona.

\begin{figure}
\begin{center}
\includegraphics[width=80mm]{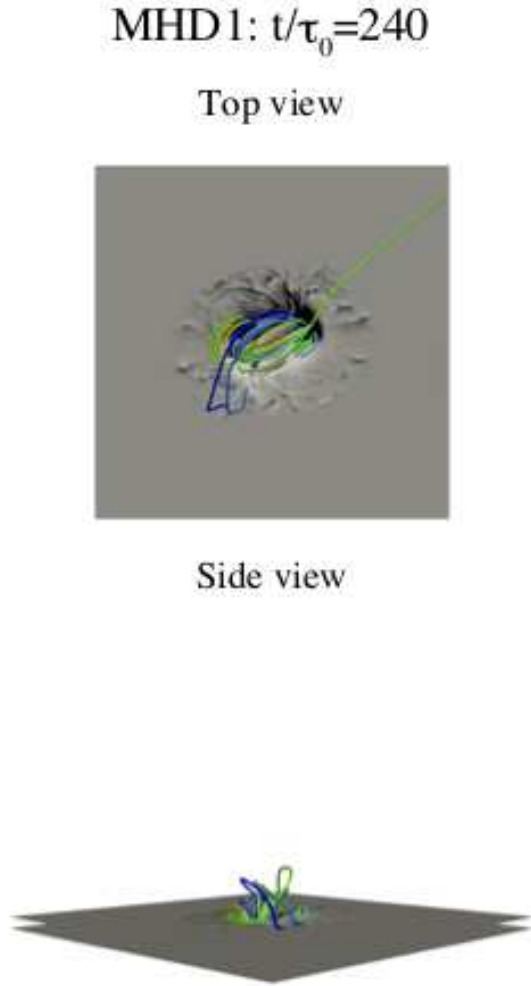}
\end{center}
\caption{Magnetic field at $t/\tau_{0}=240$ calculated by the MHD1 code using the GT slices at $z/H_{0}=10$. The two grayscale slices show $B_{z}/B_{0}$ at $z/H_{0}=0$ and 10. The seeds for integrating the field lines are identical to those in Figures \ref{fig:0240top} and \ref{fig:0240side}.\label{fig:mhd1}}
\end{figure}

The preliminary result is shown in Figure \ref{fig:mhd1}, which exhibits a confined magnetic dome, as opposed to the large-scale torus in Figures \ref{fig:0240top} and \ref{fig:0240side}. In fact, the dome structure looks much closer to the GT and other reproduced fields. The measured parameters for the new MHD1 (GT) model at $t/\tau_{0}=240$ measured above $z/H_{0}=10$ are: $E_{\rm mag}/E_{0}=591$ (132); $\Delta E_{\rm mag}/E_{0}=536$ (77); $z_{\rm apex}/H_{0}=286$ (43); and $H_{\rm R}/(B_{0}^{2}H_{0}^{4})=-3.1\times 10^{5}$ ($-0.1\times 10^{5}$), all of which are much closer to the GT values compared to the MHD1 results in Section \ref{sec:results}.

\end{document}